\documentclass[sigconf]{acmart}

\renewcommand\footnotetextcopyrightpermission[1]{}
\setcopyright{none}

\acmConference[SBES 2026]{40th Brazilian Symposium on Software Engineering}{September 8--11, 2026}{São Paulo, SP, Brazil}

\AtBeginDocument{
    
}

\usepackage{amsmath,amsfonts}
\usepackage{graphicx}
\usepackage{booktabs}
\usepackage{multirow}
\usepackage{float}
\usepackage{xcolor}
\usepackage{xspace}
\usepackage{hyperref}  %

\hypersetup{hidelinks}  %

\begin{document}

\title{Exploring the Combination of Test Generation and Test Amplification for Regression Testing}

\renewcommand{\shorttitle}{Exploring the Combination of Test Generation and Test Amplification for Regression Testing}

\author{Danilo Torquato}
\orcid{0009-0004-0157-963X}
\affiliation{%
  \institution{Universidade Federal de Pernambuco}
  \city{Recife}
  \state{PE}
  \country{Brazil}
}
\email{dtu@cin.ufpe.br}

\author{Breno Miranda}
\orcid{0000-0001-9608-9393}
\affiliation{%
  \institution{Universidade Federal de Pernambuco}
  \city{Recife}
  \state{PE}
  \country{Brazil}
}
\email{bafm@cin.ufpe.br}

\author{Antonia Bertolino}
\orcid{0000-0001-8749-1356}
\affiliation{%
  \institution{Gran Sasso Science Institute}
  \city{L'Aquila}
  \country{Italy}
}
\email{antonia.bertolino@gssi.it}

\renewcommand{\shortauthors}{Torquato et al.}

\begin{abstract}
Regression testing plays a crucial role in software engineering by ensuring that code modifications do not introduce defects into previously stable systems. However, as software projects evolve in scale and complexity, executing the full regression test suite becomes increasingly costly. To mitigate this challenge, the research community has explored diverse strategies, including test case generation and test case amplification, both of which aim to improve coverage and fault detection efficiency.
While each of these strategies has been widely investigated, in this paper we report a first preliminary empirical study on the effectiveness of combining test case generation and test case amplification. We evaluate EvoSuite (generation) and DSpot (amplification) as representatives of these strategies, both individually and in combination, using mutation score as the primary metric.
Preliminary results indicate that the combination of the two approaches achieves the highest median mutation score, surpassing individual techniques. Although improvements are not always statistically significant, the consistent trend suggests a potential synergy between generation and amplification techniques. Nevertheless, further studies are necessary to properly evaluate this trend and consolidate the initial evidence. These early findings provide initial insights for practitioners interested in regression testing optimization, while highlighting the need for more extensive investigation, particularly in large-scale and continuous integration environments.
\end{abstract}


\maketitle

\section{Introduction}

Regression testing verifies whether code modifications preserve previously stable functionalities \cite{Pan2022}. A major challenge lies in maintaining test suites that evolve alongside the system \cite{Javaria2019}. Automated techniques for test generation and amplification have emerged as promising alternatives to address this challenge.
On the one side, test case generation automates the creation of new test cases, thus relieving testers from the heavy task of manual derivation of effective regression test suites. On the other, test amplification is devoted to improve the existing test suite in terms of coverage or effectiveness.

While both techniques have been widely investigated in isolation, their potential synergies remain unexplored. Generation creates tests from scratch, exploring the code broadly, whereas amplification refines existing tests, targeting under-tested areas. These complementary strengths suggest that combining them could yield more robust test suites than either technique alone, yet no prior work has empirically evaluated this hypothesis.

While recent work demonstrated the value of orchestrating selection and prioritization strategies \cite{Fastazi}, this study examines whether the orchestration of generation and amplification yields consistent gains or depends on contextual factors.

Our results indicate that the combination generally outperforms the individual tools, particularly in more challenging scenarios. However, standalone techniques such as EvoSuite remain highly competitive, and gains from orchestration are not always statistically significant. Nevertheless, the findings reinforce the potential of integrating generation and amplification to strengthen regression test suites.

This study offers preliminary empirical insights into the benefits and limitations of orchestrating generation and amplification techniques.

\section{Background}

\subsection{Regression Testing}

Regression testing verifies that recent code changes do not adversely affect existing functionality. It re-executes previously passed test cases to ensure modifications do not introduce side effects, preserving software quality in agile environments where changes occur frequently \cite{Pan2022}.

Without adequate regression testing, even minor changes can trigger production issues, increasing maintenance costs and reducing user satisfaction \cite{MINHAS2020106254}. Regression testing is critical in CI/CD environments, where rapid releases demand stability \cite{Lai2023}.

One key challenge is scalability: as systems evolve, test suites grow, making full execution impractical \cite{Breno2018}. Test suite maintenance and flaky tests further reduce effectiveness \cite{Luo2016}. To address these challenges, techniques including test selection, prioritization, generation, and amplification have been proposed \cite{greca2023state, YooHarman2012}.

\subsection{Test Case Generation}

Test case generation is an automated technique that creates test cases based on the analysis of source code, specifications, or runtime behavior. Its primary goal is to produce tests that effectively cover the code and detect potential faults, significantly reducing the manual effort required for test creation \cite{Evosute}. This technique is particularly valuable in regression testing, where frequent test execution makes manual creation impractical.

The advantages of automated test case generation are significant. First, it reduces the manual effort needed to create and maintain test suites, making it especially suitable for large and complex systems. Second, automated tools like EvoSuite can achieve high levels of code coverage, including corner cases often overlooked in manual testing. Despite the computational effort and overhead inherent to these techniques, the ability to increase coverage and to support development teams in building effective tests makes this approach highly attractive.

Nevertheless, test case generation has limitations.
While automated tools exist that can produce tests with high coverage (see Section~\ref{sec:evosuite}), the generated cases do not always reflect real-world usage scenarios or detect complex functional faults \cite{Fan2019}. Moreover, generated tests may require manual refinement to remain relevant as software evolves, particularly in dynamic development environments. Additionally, the generation and execution process can be computationally expensive, especially for large systems or when complex algorithms are employed.

Despite these challenges, test case generation remains a promising technique for improving regression testing efficiency. Preliminary results suggest that, when combined with other approaches such as test amplification, it can further enhance regression testing effectiveness.

\subsection{Test Case Amplification}

Test case amplification is a technique that enhances existing test suites by generating additional scenarios or modifying existing ones. Unlike test generation, which creates tests from scratch, amplification focuses on improving the quality and coverage of an already existing suite \cite{Dspot_ref}.

The advantages of test case amplification are noteworthy. First, it can improve test coverage by generating additional scenarios that target untested or under-tested parts of the code. Second, amplification preserves existing tests, ensuring that previously validated functionalities remain intact. Finally, amplification is often more cost-effective than generating tests from scratch, as it leverages existing suites while focusing on areas requiring additional scrutiny.

However, test case amplification also presents limitations. Its effectiveness strongly depends on the quality of the original test suite; if the initial tests are weak, amplification may yield limited improvements in fault detection. Furthermore, amplified tests may introduce redundancy by targeting similar code paths or conditions as the original tests. Finally, amplification, particularly when using mutation-based techniques, can be computationally costly.

Despite these challenges, test case amplification represents a promising approach to improving regression testing effectiveness. When combined with test generation, it is hypothesized that such integration may provide a more comprehensive strategy for test suite optimization, leveraging the strengths of both techniques.

\subsection{Orchestration of Regression Testing Techniques}

The orchestration of regression testing techniques involves integrating two or more approaches with the aim of combining their respective strengths \cite{Fastazi}. In the context of test generation and amplification, orchestration emerges as a promising strategy for optimizing regression testing. By uniting the benefits of both techniques, it becomes possible to achieve broader code coverage, enhance fault detection, and improve testing efficiency.

The combined application of these techniques enables more effective testing of code regions that are untested or insufficiently exercised. For instance, while automated generation
contributes to creating tests with high structural coverage, amplification
reinforces and expands existing tests, focusing on critical areas that require deeper scrutiny. This integration can lead to more robust and comprehensive fault detection.

Nevertheless, orchestration also presents challenges. Integrating different tools and techniques
requires careful planning to ensure compatibility and seamless operation. Moreover, the initial configuration and execution of combined techniques may introduce additional overhead, particularly in terms of computational resources. Ensuring the quality of generated and amplified tests is also critical, as poorly constructed tests may lead to false positives, redundancy, or undetected faults.

Despite these obstacles, orchestrating generation and amplification techniques holds significant potential for improving regression testing. By overcoming technical and operational challenges, software engineers can harness the combined strengths of these approaches to achieve more effective and efficient testing outcomes \cite{Fastazi}.

\section{Combining Test Generation and Amplification Techniques}

This section describes the state-of-the-art tools and evaluation metrics used in the study. In particular, we evaluate EvoSuite (test generation) and DSpot (test amplification), as well as their combination (EvoSuite + DSpot).

\subsection{EvoSuite: Search-Based Test Generation}
\label{sec:evosuite}

Among the most widely used tools in academia for test case generation is EvoSuite. EvoSuite employs search-based software testing techniques, such as genetic algorithms, to generate test cases that maximize code coverage. By analyzing the software’s bytecode, EvoSuite evolves test cases that cover branches, statements, and other structural elements. This tool is particularly effective in generating high-coverage tests, making it a popular choice for regression testing in Java projects \cite{Evosuite2020}. Studies have shown that EvoSuite can produce high-quality test suites that are effective in fault detection while maintaining reasonable execution times \cite{Evosute}, even in industrial environments \cite{Almasi2017}.

EvoSuite offers several advantages. First, it achieves high levels of branch and statement coverage, often discovering edge cases overlooked by manual testing. Second, it integrates seamlessly with build systems such as Maven and Gradle, making it practical for continuous integration pipelines. Finally, EvoSuite automatically generates assertions to verify the correctness of produced tests, reducing the need for manual intervention \cite{10.1145/2025113.2025179}. However, generated tests may sometimes lack real-world relevance or fail to detect complex functional faults, and the process can be computationally expensive for large systems.

\subsection{DSpot: Mutation- and Coverage-Based Test Amplification}

A prominent tool for test amplification is DSpot \cite{Dspot_ref}.
Unlike test generation tools that create tests from scratch, DSpot expands existing tests by applying amplification operators and analyzing coverage metrics. Its objective is to increase the effectiveness of regression testing by promoting greater diversity and depth in the test cases, with special attention to under-tested or fault-prone areas of the code.

Its approach relies on amplification operators that modify and extend tests—for instance, by altering literal values, inserting new assertions, and exploring alternative execution paths. Furthermore, DSpot leverages coverage and behavioral metrics to guide modifications, aiming to increase the suite’s sensitivity to subtle faults. However, its effectiveness depends on the quality of the original suite: if the baseline tests are weak, amplification may yield only modest improvements. Another consideration is that amplification can be computationally costly, particularly in large-scale systems.

\subsection{Combining Generation with EvoSuite and Amplification with DSpot}

The combination of EvoSuite and DSpot integrates evolutionary search-based automated test generation with amplification of existing tests via amplification operators. EvoSuite generates an initial suite with high structural coverage, while DSpot operates exclusively on the developer's test suite, applying amplification operators to strengthen assertions, explore input variations, and target under-tested code regions. The resulting test suites from both tools were merged, and compilation errors were resolved by excluding problematic tests until all suites compiled successfully. This integration aims to analyze whether the two techniques are complementary, i.e., whether each tool kills mutants that the other cannot.

\section{Evaluation Methodology}

\subsection{Research Question and Evaluation Design}

This section presents the research question and the experimental design adopted to investigate the effectiveness of test generation and amplification techniques, both individually and in combination. It also describes the metrics used throughout the study.

The study is guided by the following research question (RQ):

\begin{itemize}
    \item \textbf{RQ: How do the test generation tool EvoSuite, the test amplification tool DSpot, and their combination perform in terms of effectiveness, considering the Mutation Score as the evaluation metric?}
\end{itemize}

This question allows for a detailed analysis of the individual and combined performance of the tools, enabling the identification of their strengths and limitations.

The effectiveness of the evaluated combinations was assessed using the mutation score metric, following the methodology from the study conducted by Roslan, Rojas, and McMinn, who
empirically compared EvoSuiteAmp, a modified version of EvoSuite for test amplification, and DSpot individually, evaluating their impact on the mutation score of test suites \cite{10.1007/978-3-031-21251-2_2}.

\subsection{Mutation Score}

Traditional structural coverage metrics, such as line and branch coverage, do not necessarily correlate with the actual fault detection capability of a test suite \cite{Inozemtseva2014}. Previous studies have shown that code coverage alone is not strongly associated with increased test effectiveness, reinforcing the need for stronger evaluation metrics \cite{Inozemtseva2014}. In this context, mutation score has been widely adopted as an indicator of test effectiveness, since it evaluates the ability of a test suite to detect artificially injected faults.

Mutation testing helps reveal weaknesses in a test suite by introducing small artificial faults into the code and evaluating whether the tests detect them. Consequently, mutation score provides a more fault-oriented assessment of test quality than traditional structural coverage metrics \cite{10.1145/3635713}. Mutation score is computed as the percentage of mutants killed by a test suite.

High mutation scores generally indicate that a test suite is effective at detecting a broader range of potential faults. However, mutation analysis can be computationally expensive, especially for large and complex systems \cite{10.1007/s10586-023-04074-y}.

In this study, mutation score was used to compare the effectiveness of the tested combination (EvoSuite + DSpot) in fault detection. By measuring the percentage of mutants killed by the combination, it is possible to assess their strengths and limitations in terms of fault detection. Mutation score is calculated using the following formula:

\begin{equation}
\text{Mutation Score (\%)} =
\frac{\text{Number of Killed Mutants}}{\text{Total Number of Mutants}} \times 100
\end{equation}

A higher score indicates a more effective test suite, capable of detecting a greater number of potential faults introduced through code mutations \cite{PAPADAKIS2019275}.

The design and execution of the experiment were based on the methodology described on Fastazi \cite{Fastazi}. The experiments aimed to compare four possible test arrangements across the following group:

\begin{itemize}
\item \textbf{Group}: The developers' original test suite, the suite generated by EvoSuite, the suite amplified by DSpot, and the orchestration of EvoSuite and DSpot.

\end{itemize}

The developer's original test suite serves as the baseline to assess the effectiveness of automated approaches.

\subsection{Experimental Design and Execution}

The experiment used twelve projects available in the Defects4J repository \cite{just2014defects4j}, from which ten Maven-based projects were selected. Defects4J contains multiple versions of Java-based open-source software projects, each consisting of a commit containing a bug, the commit that fixed the bug, and metadata such as related files and tests that would detect the bug. The complete replication package, including all generated and amplified test suites and mutation analysis results, is publicly available \cite{zenodo2026}.

For this experiment, the most recent version of each project was initially used to evaluate the tools. After analyzing the latest version of each project, we proceeded to analyze the last five versions of each project that contained classes meeting all the established criteria.

\begin{itemize}
\item \textbf{EvoSuite}: Version 1.2.0, using the standalone JAR executable available at \url{https://www.evosuite.org/}.
\item \textbf{DSpot}: Version 3.2.0, used as a plugin for Maven build systems, available at \url{https://github.com/STAMP-project/dspot}.
\end{itemize}

To obtain the Mutation Score, the Major tool integrated with Defects4J was used, as described in \cite{10.1007/978-3-031-21251-2_2}. Major generates mutants by applying mutation operators to the source code, and the test suites are executed against these mutants to measure their effectiveness. To account for the non-deterministic behavior of EvoSuite, DSpot, and Major, outputs were generated 30 times to reduce potential noise in the data.

During the study, rules were adopted based on the guidelines established by Rojas et al. \cite{10.1007/978-3-031-21251-2_2}. First, only the most recent version of each Defects4J project was considered. After analyzing these latest versions, we selected only those projects where both EvoSuite and DSpot successfully generated or amplified test cases for at least one class. Finally, we narrowed our analysis to include only the classes in which both tools improved the developer's mutation score. This filtering allowed us to assess complementarity, as analyzing classes where one tool fails would not permit such assessment. The final dataset comprised six Defects4J projects.

Moreover, since DSpot amplifies test cases, it sometimes relies on auxiliary files written by developers, such as utility classes. To avoid compilation errors, it was ensured that the amplified test suites included all necessary imports for these dependencies.

Furthermore, the evaluation focused on analyzing the complementary fault detection capability provided by generation and amplification techniques. the study investigated how each approach contributed to mutation detection by identifying mutants uniquely killed by EvoSuite, uniquely covered through DSpot amplification, and mutants detected by both techniques. This perspective allowed the analysis to highlight the complementary behavior between the approaches and to better characterize the potential benefits of combining generation and amplification techniques in regression testing.

\section{Results}
\label{sec:results}

This section presents and analyzes the results obtained, focusing on the effectiveness and efficiency of the tested combinations of test generation and amplification techniques. We start with a visual analysis using boxplots, followed by a statistical analysis to assess the significance of the findings. Finally, the implications of the results are discussed.

To evaluate the effectiveness (RQ) of the tested combinations, a visual analysis using boxplots was first conducted. The plot displays separate boxplot for each individual tool and for the combination, allowing a direct comparison of the achieved mutation score.

\begin{figure}[t]
    \centering
    \includegraphics[width=0.7\linewidth]{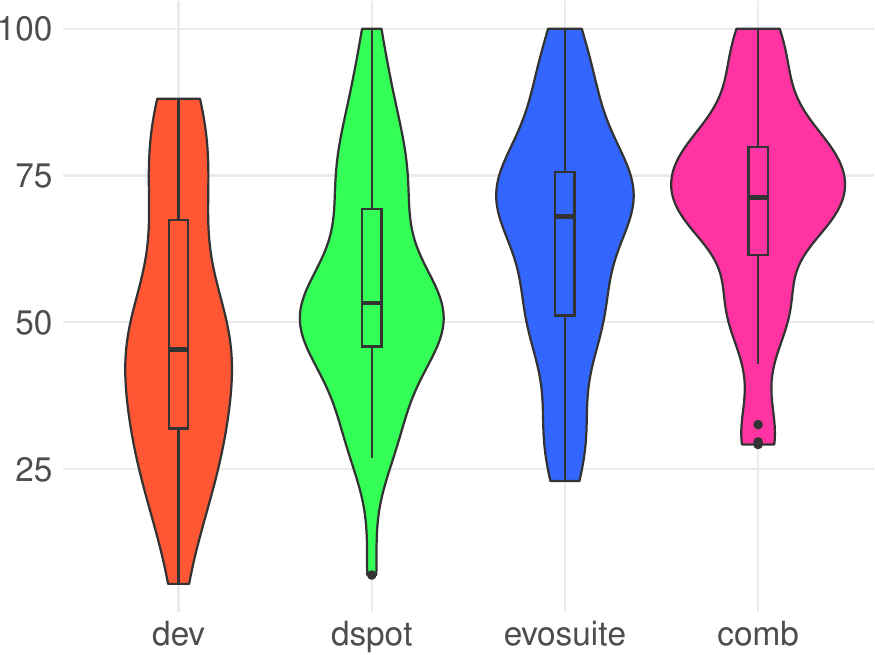}
    \caption{Mutation score for Dev, DSpot, EvoSuite, and their combination.}
    \Description{Boxplot comparing mutation coverage of DSpot, EvoSuite, and their combination. The combination shows the highest median coverage.}
    \label{fig:Evo_dspot_boxplot}
\end{figure}

Figure~\ref{fig:Evo_dspot_boxplot} presents the distribution of mutation score for the EvoSuite and DSpot combination, along with the individual tools. The boxplot provides a clear comparison of their performance.

The EvoSuite + DSpot combination achieves a higher median mutation score than either individual tool. This demonstrates a trend of benefits from combining EvoSuite’s high-coverage test generation with DSpot’s mutation-based amplification. EvoSuite, as a standalone tool, performs well, often outperforming DSpot in coverage, highlighting the effectiveness of evolutionary algorithms in achieving high code coverage. DSpot, on the other hand, effectively enhances the test suite by targeting under-tested or fault-prone areas, complementing EvoSuite’s strengths. The combination also exhibits relatively consistent performance across projects.

The visual analysis reveals several important trends. First, the combination outperforms the individual components in terms of mutation score, indicating a positive trend of combining test generation and amplification techniques. This can be attributed to the complementary nature of the tools: EvoSuite contributes with high-coverage test generation and DSpot performs mutation-based amplification, making tests more effective at detecting faults. Additionally, performance variability across projects highlights the importance of adapting test strategy choices to the specific characteristics of the software under test.

Table~\ref{tab:all_results_part1} presents detailed results for each tool and combination, including mutants generated, mutation score, and killed mutants. The developer test suite (Dev) shows that although some classes achieve good mutation scores (e.g., OptionGroup and QuotedPrintable), overall coverage tends to be limited. These data reinforce the need for automated approaches to enhance test effectiveness.

Regarding test amplification with DSpot, several classes show a significant increase in killed mutants compared to Dev, with notable improvements in CSVFormat, ZipFile, and Tag. However, improvement is not substantial across all classes, highlighting that amplification effectiveness depends on the quality of the original suite.

EvoSuite demonstrates strong performance across multiple classes, achieving full coverage in OptionGroup and high scores in CSVRecord, TimedSemaphore, and HashCodeBuilder. Nevertheless, some classes (e.g., ZipFile) still exhibit limitations, possibly due to internal complexity or external dependencies.

The combination of both techniques shows consistent increase in mutation coverage compared to individual tools. Classes such as CSVFormat, ZipFile, and Attributes present significant improvements. This highlights the potential of orchestration to produce more robust test suites, reinforcing the gains observed in the statistical analysis.

\begin{table}[t]
\centering
\scriptsize
\setlength{\tabcolsep}{4pt}
\caption{Results: Mutation Score (\%) and Killed Mutants}
\label{tab:all_results_part1}
\begin{tabular*}{\linewidth}{@{\extracolsep{\fill}} p{1.5cm} c c c c c c c c c}
\toprule
\textbf{Class} & \textbf{Mut} & \multicolumn{2}{c}{\textbf{Dev}} & \multicolumn{2}{c}{\textbf{DSpot}} & \multicolumn{2}{c}{\textbf{EvoSuite}} & \multicolumn{2}{c}{\textbf{Comb}} \\
\cline{3-10}
 & & \% & K & \% & K & \% & K & \% & K \\
\midrule
\textbf{CLI – 40} & & & & & & & & & \\
OptionGroup & 23 & 86.96 & 20 & 91.30 & 1 & 100.00 & 3 & 100.00 & 3 \\
Option & 215 & 43.72 & 94 & 60.00 & 35 & 73.95 & 65 & 79.53 & 77 \\
Options & 45 & 33.33 & 15 & 55.56 & 10 & 51.11 & 8 & 66.67 & 15 \\
HelpFormatter & 356 & 69.66 & 248 & 69.94 & 1 & 77.81 & 29 & 77.81 & 29 \\
OptionBuilder & 54 & 44.44 & 24 & 46.30 & 1 & 66.67 & 12 & 68.52 & 13 \\
PatternOptBuilder & 95 & 73.68 & 70 & 74.74 & 1 & 81.05 & 7 & 82.11 & 8 \\
\midrule
\textbf{CLI – 38} & & & & & & & & & \\
Option & 215 & 38.60 & 83 & 57.20 & 40 & 72.55 & 73 & 77.67 & 84 \\
\midrule
\textbf{Codec – 18} & & & & & & & & & \\
Base64 & 728 & 68.41 & 498 & 68.68 & 2 & 71.84 & 25 & 72.12 & 27 \\
HmacUtils & 37 & 51.35 & 19 & 54.05 & 1 & 67.57 & 6 & 67.57 & 6 \\
QuotedPrintable & 224 & 81.25 & 182 & 82.14 & 2 & 89.73 & 19 & 89.73 & 19 \\
\midrule
\textbf{Compress – 47} & & & & & & & & & \\
DumpArchiveEntry & 331 & 5.44 & 18 & 6.95 & 5 & 29.31 & 79 & 29.61 & 80 \\
TarArchiveEntry & 505 & 26.93 & 136 & 31.88 & 25 & 50.30 & 118 & 50.30 & 118 \\
ZipFile & 431 & 21.11 & 91 & 53.13 & 138 & 22.97 & 8 & 53.60 & 140 \\
BlockSort & 1953 & 30.47 & 595 & 30.52 & 1 & 32.51 & 40 & 32.57 & 41 \\
\midrule
\textbf{Compress – 46} & & & & & & & & & \\
ZipArchiveIS & 925 & 26.70 & 247 & 6.95 & 2 & 28.97 & 21 & 29.18 & 23 \\
StreamCompressor & 115 & 36.52 & 42 & 31.88 & 7 & 49.56 & 15 & 49.56 & 15 \\
TarArchiveIS & 333 & 54.95 & 183 & 53.13 & 1 & 57.95 & 10 & 58.25 & 11 \\
ZipFile & 428 & 50.70 & 217 & 30.52 & 4 & 51.16 & 2 & 52.10 & 6 \\
\midrule
\textbf{Csv – 16} & & & & & & & & & \\
CSVFormat & 587 & 31.35 & 184 & 53.32 & 129 & 48.38 & 100 & 67.46 & 212 \\
CSVRecord & 31 & 74.19 & 23 & 77.42 & 1 & 96.77 & 7 & 96.77 & 7 \\
\midrule
\textbf{Jsoup – 93} & & & & & & & & & \\
Element & 518 & 66.41 & 344 & 67.76 & 7 & 74.32 & 41 & 75.48 & 47 \\
Node & 210 & 44.76 & 94 & 48.57 & 8 & 68.57 & 50 & 69.05 & 51 \\
Tag & 226 & 17.26 & 39 & 76.11 & 133 & 32.30 & 34 & 80.09 & 142 \\
Attributes & 271 & 32.47 & 88 & 46.49 & 38 & 67.90 & 96 & 76.75 & 120 \\
Attribute & 105 & 19.05 & 20 & 38.10 & 20 & 51.43 & 34 & 64.76 & 48 \\
TextNode & 92 & 34.78 & 32 & 55.43 & 19 & 71.74 & 34 & 77.17 & 39 \\
CharacterReader & 600 & 43.83 & 263 & 44.67 & 5 & 76.00 & 193 & 76.00 & 193 \\
TokenQueue & 278 & 47.48 & 132 & 47.84 & 1 & 75.18 & 77 & 75.18 & 77 \\
\midrule
\textbf{Jsoup – 88} & & & & & & & & & \\
Attribute & 104 & 15.38 & 16 & 34.61 & 20 & 60.57 & 47 & 71.15 & 58 \\
\midrule
\textbf{Jsoup – 91} & & & & & & & & & \\
TokenQueue & 278 & 46.40 & 129 & 46.76 & 1 & 67.98 & 60 & 67.98 & 60 \\
\midrule
\textbf{Lang – 28} & & & & & & & & & \\
StringUtils & 2914 & 45.33 & 1321 & 45.44 & 3 & 67.30 & 640 & 67.40 & 643 \\
HashCodeBuilder & 328 & 81.40 & 267 & 83.84 & 8 & 90.55 & 30 & 90.85 & 31 \\
TimedSemaphore & 67 & 88.06 & 59 & 100.00 & 8 & 92.54 & 3 & 100.00 & 8 \\
Fraction & 889 & 79.64 & 708 & 80.09 & 4 & 83.13 & 31 & 83.46 & 34 \\
FieldUtils & 83 & 60.24 & 50 & 68.67 & 7 & 74.70 & 12 & 83.13 & 19 \\
StrBuilder & 1832 & 49.89 & 914 & 50.22 & 6 & 71.18 & 390 & 71.28 & 392 \\
StrSubstitutor & 217 & 87.56 & 190 & 91.24 & 8 & 88.94 & 3 & 93.63 & 11 \\
\bottomrule
\end{tabular*}
\vspace{1mm}
\begin{flushleft}
\footnotesize
\textbf{Legend:} Mut = Generated Mutants; \% = Mutation Score; K = Killed Mutants; DSpot, EvoSuite, and Comb values indicate the increase in killed mutants.
\end{flushleft}
\end{table}

\subsection{Statistical Analysis}

\label{subsec:statistical}

To further assess the effectiveness of the tested combination, a statistical analysis was conducted using the Kruskal-Wallis test, followed by pairwise comparisons through the Wilcoxon test with Bonferroni correction for significance adjustment. The Kruskal-Wallis test was employed to determine whether there were statistically significant differences in mutation coverage among the evaluated approaches. The null hypothesis assumed no differences between groups. The test yielded a statistically significant result (p < 0.05), indicating that at least one group differs from the others. The results of these comparisons are summarized in the statistical groups below.

The statistical analysis grouped the evaluated approaches based on their mutation coverage, revealing significant differences among them. The EvoSuite + DSpot combination (Comb) showed the highest median mutation coverage (71.28) and was allocated to Group A. This indicates that this combination is statistically superior to the lower groups (BC and C). EvoSuite alone, with a median of 67.98, was classified in Group AB.
DSpot alone obtained a median of 53.32 and was included in Group BC, indicating lower performance than EvoSuite but still higher than the baseline approach (Dev). The baseline, with a median of 45.33, was assigned to Group C, being statistically inferior to all other approaches.

These results provide several important insights. First, although the EvoSuite + DSpot combination is not statistically superior to EvoSuite alone, it achieves the highest median mutation coverage and belongs to the highest statistical group (Group A). This points to a strong trend of superior performance, even if the difference did not reach statistical significance. EvoSuite demonstrates solid performance, with results comparable to the combination, highlighting the effectiveness of evolutionary search-based test generation as a standalone approach. However, considering the combination with DSpot suggests additional gains from amplification techniques. The baseline approach (Dev) is generally inferior to the automated approaches, reinforcing the importance of test generation and amplification techniques to enhance coverage.

The performance of the tools and their combinations varies across the analyzed projects, suggesting that the choice of testing strategy should be carefully adapted to the specific characteristics of the software under test. This variability underscores the need for context-sensitive decisions in regression testing processes.

In this regard, the results of this study have practical implications for developers. Initially, it is recommended to consider combined approaches, such as integrating EvoSuite with DSpot, to enhance the effectiveness of regression testing. Although the gains are not always statistically significant, there is a clear trend of improved outcomes. Furthermore, the variation in performance across different projects emphasizes the importance of adapting testing strategies to the specific requirements of each context. In some cases, using EvoSuite alone may suffice, but in others, combining it with amplification techniques, such as those offered by DSpot, can provide additional benefits and greater robustness in fault detection.

The statistical analysis of the mutation scores is summarized in Table \ref{tab:statistical}.

\begin{table}[t]
\centering
\caption{Mutation Score Comparison}
\label{tab:statistical}
\small
\setlength{\tabcolsep}{4pt}
\begin{tabular*}{\linewidth}{@{\extracolsep{\fill}} l c c c c}
\toprule
Metric & Dev & DSpot & EvoSuite & Comb \\
\midrule
med   & 45.33 & 53.32 & 67.98 & 71.28 \\
sd    & 22.35 & 20.00 & 19.67 & 17.82 \\
group & c & bc & ab & a \\
\bottomrule
\end{tabular*}
\end{table}

\section{Lessons Learned}

Visual inspection reveals that the combination's benefit depends on the initial quality of the developer's test suite. Using the median developer mutation score (45.33\%) as a split point, we analyzed two groups: low initial coverage (below the median) and high initial coverage (at or above the median).

\begin{figure}[htbp]
    \centering
    \includegraphics[width=0.32\textwidth]{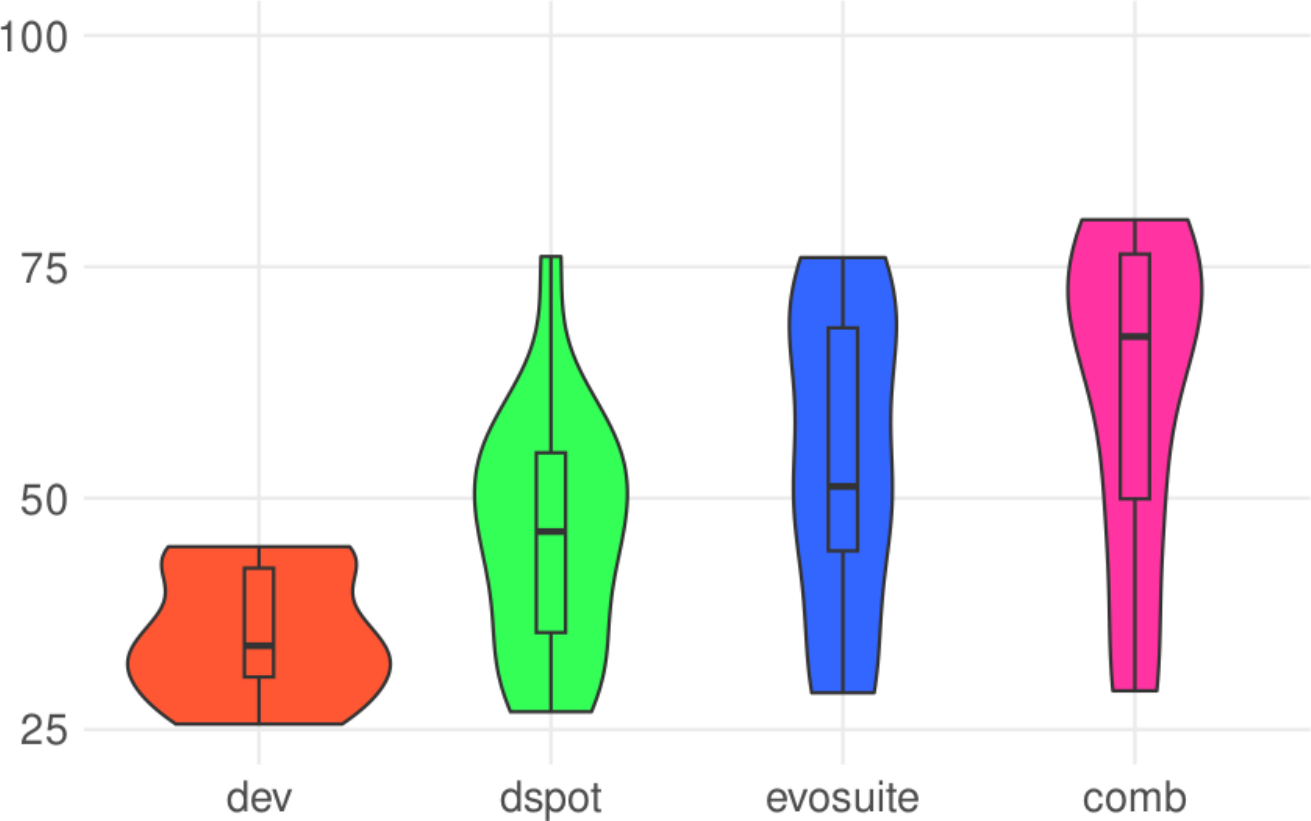}
    \hspace{0.05\textwidth}
    \includegraphics[width=0.32\textwidth]{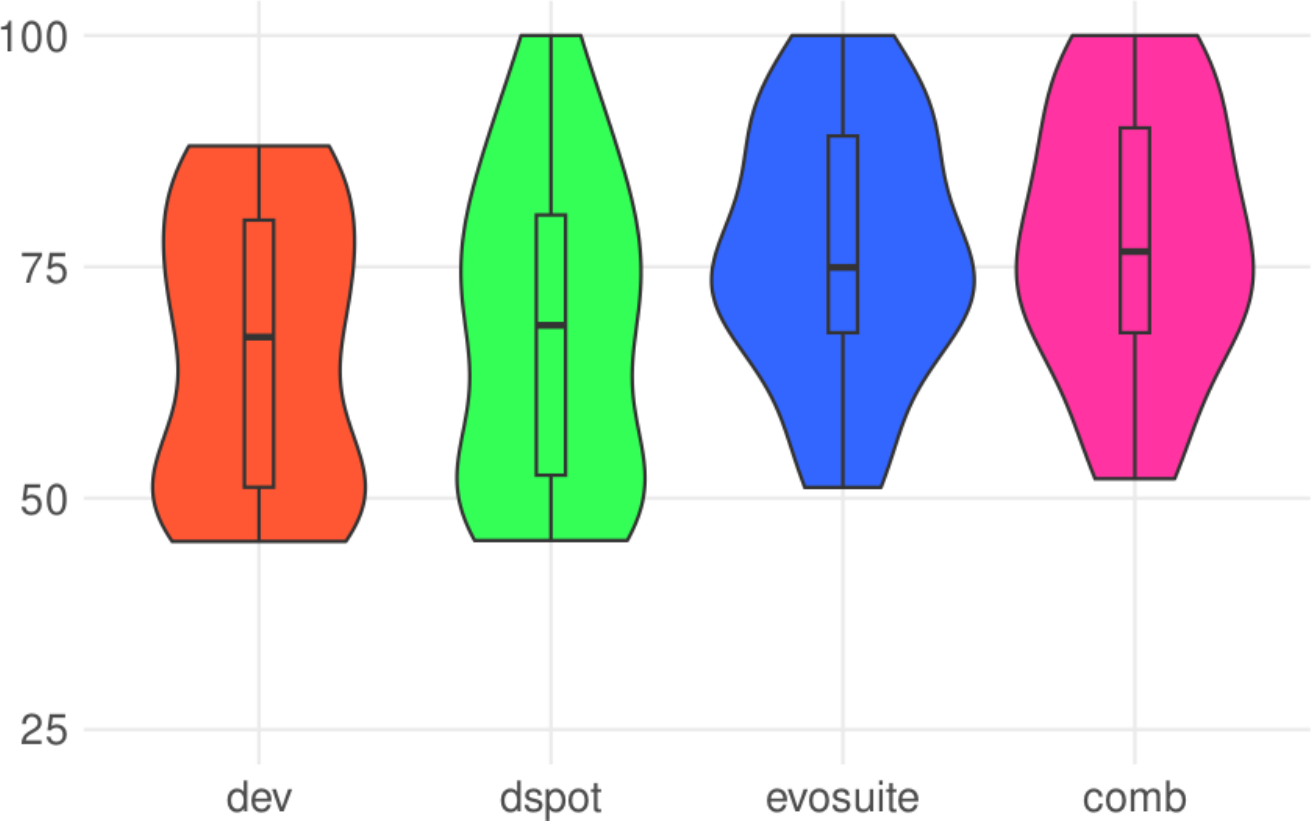}
    \caption{Mutation coverage: low initial coverage (top) vs high initial coverage (bottom)}
    \label{fig:coverage_thresholds}
\end{figure}

For low coverage classes, the combination achieves 67.46\% versus EvoSuite's 51.11\% (gap of 16.35 points). For high coverage classes, the gap narrows to 1.71 points. This suggests orchestration provides greater returns where tests are weakest. However, pairwise Wilcoxon tests did not reach statistical significance, indicating that larger samples are needed for confirmation.

This pattern suggests that the benefit of combining generation and amplification is not universal but structural: DSpot contributes meaningfully only where EvoSuite's search-based approach fails to reach. Identifying such contexts a priori, before committing to the overhead of orchestration, remains an open and compelling research challenge.

\section{Threats to Validity}

The experiments used projects from Defects4J, which may not represent the full diversity of software systems. Only the most recent version of each project was analyzed, limiting the ability to observe behavior across software evolution.

Mutation coverage (Major tool) captures only one dimension of testing quality, ignoring aspects such as test readability, maintainability, and execution cost.

EvoSuite, DSpot, and Major are non-deterministic. To mitigate this, each configuration was executed 30 times, though some residual variability may remain.

It is assumed tests are deterministic. No flaky behavior was observed, but environmental conditions could influence results.

Specific versions and default configurations of EvoSuite and DSpot were used. Different configurations or future versions could produce different results.

Finally, the number of analyzed classes may limit statistical power for detecting small differences. Larger samples could reveal additional effects.

\section{Related Work}

Most studies on automated regression testing focus on test generation or test amplification in isolation, evaluating tools such as EvoSuite and DSpot independently. However, studies analyzing the combination of these techniques are still scarce.

EvoSuite employs evolutionary search to generate unit tests focused on branch, line, and mutation coverage. Empirical evaluations show that EvoSuite achieves significant results, as evidenced in SBST 2021, where it achieved the highest score among all evaluated tools \cite{9476230}. Recent studies also highlight advances in EvoSuite, such as more sophisticated search algorithms and improved stability \cite{Robert2021}.

Randoop generates tests through random execution of available methods. Although simpler and faster, it may produce tests with lower precision and more redundant test cases, with effectiveness varying depending on the application structure \cite{Bastos_Almeida_2024}.

An evaluation with software engineers identified a preference for tools providing better oracle assertiveness and greater control over generated tests \cite{Ali2010}.

In test amplification, DSpot starts from manually written developer tests and applies Assertion Amplification and Input Amplification. Evaluated on open-source Java projects, DSpot increased mutation scores and generated improvements accepted by developers \cite{Dspot_ref}.

A systematic literature review classified amplification approaches into four categories: modification of test code, addition of test variants, modification of test execution, and synthesis of new tests guided by code changes \cite{DANGLOT2019110398}.

Despite these advances, most existing studies still evaluate test generation and test amplification independently. There is limited empirical evidence regarding how these techniques interact when applied together, particularly considering complementary fault detection capabilities. Our study addresses this gap by comparing individual and combined approaches.

\section{Conclusions}

This preliminary study investigates whether combining EvoSuite and DSpot improves regression testing effectiveness. Using mutation coverage on the Defects4J projects, we found that:

Combination achieves highest median coverage: The combination reached 71.28\%, compared to EvoSuite (67.98\%), DSpot (53.32\%), and developer suites (45.33\%).

Benefits are context-dependent: Orchestration provides larger gains for classes with weak initial test suites (16.35 point improvement) than for well-tested ones (1.71 points).

Synergy is context-dependent: DSpot becomes redundant where EvoSuite already covers the mutant space effectively. The largest gains emerge precisely where EvoSuite faces structural limitations, suggesting that context-aware orchestration, rather than blind combination, is the most promising research direction.

These findings are preliminary. Further studies with larger samples are needed to confirm statistical significance. For practitioners: consider prioritizing the combined approach (EvoSuite + DSpot) on modules with historically low test coverage for maximum gain. For researchers, orchestration of generation and amplification is a promising direction requiring more extensive validation.

\section*{Artifact Availability}
The test suites, mutation analysis results, and supporting resources are available in our replication package~\cite{zenodo2026}.

\section*{Acknowledgements}
This work was partially supported by CNPq grants 315765/2023-2 and 408651/2023-7, and FACEPE grant IBPG-1214-1.03/22.

\balance
\bibliographystyle{ACM-Reference-Format}
\bibliography{biblio}

\end{document}